\documentclass{aa}
\usepackage{graphicx}
\usepackage{natbib}
\usepackage{color}
\bibpunct{(}{)}{;}{a}{}{,} 
\begin{document}

\def\ms{M$_{\odot}$}
\def\zs{Z$_{\odot}$}
\def\mga{$^{24}$Mg}
\def\mgb{$^{25}$Mg}
\def\mgc{$^{26}$Mg}

\title{Light nuclei in galactic globular clusters : constraints on the self-enrichment scenario from nucleosynthesis}

\author{ Nikos Prantzos\inst{1}, Corinne Charbonnel\inst{2,3} and Christian Iliadis\inst{4}
        }

\authorrunning{N. Prantzos, C. Charbonnel and C. Iliadis}
 
\titlerunning{Nucleosynthesis in globular clusters}


\institute{ Institut d'Astrophysique de Paris, UMR7095 CNRS, Univ.P. \& M.Curie,  98bis Bd. Arago, 75104 Paris, France, 
                \email{prantzos@iap.fr}
\and 
Geneva Observatory, University of Geneva, chemin des Maillettes 51, 1290 Sauverny, Switzerland
\email{Corinne.Charbonnel@obs.unige.ch}
\and
Laboratoire d'Astrophysique de Toulouse et Tarbes, CNRS UMR 5572, OMP, 14, av.E.Belin, 31400 Toulouse, France
\and
Dpt. of Physics and Astronomy, University of North Carolina, Chapel Hill, NC 2759-3255, USA
\email{iliadis@unc.edu}            }
\date{Submitted January , 2007}

\abstract{}{Hydrogen-burning is the root cause of the 
star-to-star abundance variations 
of light nuclei in Galactic globular clusters (GC). In the present work we 
constrain the physical conditions that gave rise to the observed abundance 
patterns of Li, C, N, O, Na, Mg, Al, as well as Mg isotopes in the typical case
of NGC~6752.}
{We perform nucleosynthesis calculations at constant temperature, adopting 
realistic initial abundances for the proto-cluster gas. We use a 
detailed nuclear reaction network and state-of-the-art nuclear reaction 
rates.}
{Although simplistic, our analysis provides original 
results and new constraints on the self-enrichment scenario for GCs. 
Our parametric calculations allow us to determine a narrow range of temperature 
where the observed extreme abundances of all light elements and isotopes 
in NGC~6752 are nicely reproduced simultaneously.
This agreement is obtained after mixing of the H-processed material with 
$\sim 30 \%$ of unprocessed gas. 
We show that the observed C-N, O-Na, Mg-Al, Li-Na and F-Na anticorrelations 
as well as the behaviour of the Mg isotopes can be recovered by assuming mixing 
with various dilution factors. Li production by the 
stars that build up the other abundance anomalies is not mandatory in the case 
of NGC~6752.}
{Observations of O, Na, Mg and Al constrain the temperature range for H-burning; such temperatures are encountered in
the two main candidate ``polluters'' proposed  for GCs, namely massive AGBs and the most massive main-sequence stars.
Furthermore, observations require dilution of H-burning processed material with pristine one. 
They provide no clue, however, as to the nature of the 
unprocessed material required for mixing. The complementary observations of 
the fragile Li and F clearly point to ISM origin for the mixed material.  }

\keywords{Stars: abundances; Galaxy: abundances, globular clusters:general, 
globular clusters: individual: NGC~6752}

\maketitle

\section{Introduction}

Clues on the early 
evolution of Galactic globular clusters (hereafter GCs) may be found in the 
peculiar abundance patterns of their long-lived low-mass stars that have been 
intensively studied since the early seventies\footnote{We refer to Gratton et 
al.~(2004) and Sneden~(2005) for extensive reviews of the very broad literature 
related to the chemical properties of GCs; see also references in 
Prantzos \& Charbonnel~(2006, hereafter PC06).}.
Inside any well-studied individual GC these stars present indeed large 
abundance variations of the light nuclei from C to Al, although their content 
in heavier elements 
(i.e., Fe-group, $\alpha$-elements etc.) appears to be fairly 
homogeneous\footnote{This is not the case of $\omega$ Cen which is not a typical GC 
but may be considered as the surviving remnant of a larger system 
(e.g. Smith 2004 and references therein).}.

H-burning through the CNO-cycle and the NeNa- and MgAl-chains has 
been identified as the root cause of the chemical inhomogeneities of light nuclei 
in GCs (Kudryashov \& Tutukov 1988; Denisenkov \& Denisenkova 1989, 1990; 
Langer et al. 1993; Langer \& Hoffman 1995).
High-resolution spectroscopic analyses performed with very large 
telescopes have revealed that the O-Na and Mg-Al anticorrelations discovered 
first among evolved GC stars on the red giant branch are also 
exhibited by subgiant and turnoff stars\footnote{Note that the corresponding 
large star-to-star abundance variations have never been observed among field 
stars with comparable metallicity [Fe/H], age, and evolutionary status 
(Hanson et al. 1998; Gratton et al. 2000; Palacios et al. 2002; Mishenina et al. 2006). 
Environment, therefore, seems to have a crucial influence on the occurrence 
of these chemical patterns.}. 
This extension of the abundance anomalies to unevolved or scarcely evolved 
GC stars was known for long regarding the C-N anticorrelation.
It is crucial, because it provides compelling evidence that these patterns can not 
be due to in situ nuclear reactions\footnote{In  low-mass 
turnoff stars the temperature is not high enough to allow the proton-captures 
reactions in the NeNa- and MgAl-chains as will be discussed later on in more details.}, 
but that they were already present in the material out of which 
the observed stars formed. 
This is the so-called {\it self-enrichment} scenario that 
implies (1) pollution of the intracluster gas by H-processed material ejected 
by relatively massive and rapidly evolving stars in the early stages of the cluster 
life (Cottrell \& Da Costa 1981; Smith \& Norris 1982) and 
(2) formation of a second generation of stars by 
the ejecta or by a mixture of the ejecta with pristine material. 
Observations suggest  that all metals heavier than Si were already implanted in 
the proto-cluster gas (i.e., their present abundance is pristine)
and that inside the GC itself only the abundances of elements up to Al have been modified 
with respect to the initial cluster composition (which is the composition of 
contemporary halo field stars, see PC06 for more details).

The stellar sources responsible for the GC chemical anomalies have not been
indubitably identified yet.
As noticed by several authors, understanding the most extreme 
abundances exhibited by stars in GCs is a major key to understand the 
overall abundance patterns. Several important theoretical efforts have been made 
towards that goal, with custom-made stellar evolution models of various degrees 
of sophistication. 
Competing nucleosynthesic sites today 
are massive AGB stars on one hand (Ventura et al. 2001; D'Antona et al. 2002, Denissenkov and  Herwig
2003), 
and fast rotating massive stars on the other (PC06; Decressin et al. 2007). 
Both candidate polluters present advantages and problems, not only from the point 
of view of nucleosynthesis, but also regarding for instance the required numbers of progenitor stars
(i.e. the initial mass function of the 1st generation)
or the existence of a triggering process to ignite 
the formation of the chemically peculiar low-mass stars of the 2nd generation (see PC06 for more details). 
In both cases the complexity of the involved physical phenomena certainly 
requires deeper insight. 

In view of the important uncertainties still affecting  stellar nucleosynthesis models 
we decided to tackle the problem from a different angle. 
In the present work we discuss the case of NGC~6752, which is one of  the 
best studied GC regarding its chemical properties; these properties  are presented in \S~2. 
In \S~3 we aim at finding whether, starting with realistic
initial abundances for the protocluster gas, one may reproduce the observed 
extreme abundances of all the light nuclei for some narrow range of temperature
and timescale of H-burning. For that purpose we perform nucleosynthesis 
calculations at constant temperature, which is a satisfactory approximation for the quiescent stellar 
burning stages that we have in mind (i.e. central H-burning for massive, mass losing, stars and hot-bottom burning in massive AGB stars). We then explore the dilution factors 
between H-processed and unprocessed material that are required to reproduce
the C-N, O-Na, Mg-Al anticorrelations as well as the behaviour of the Mg 
isotopes in NGC~6752. In \S~4 we discuss the case of the anticorrelations 
observed between Na and the light nuclei Li and F.  
Based on the strong constraints on temperature that we get from our nucleosynthesis calculations,
we discuss in \S~5 the corresponding astrophysical sites. 
In \S~6 we summarize our results.

\section{Abundances in NGC~6752}

NGC~6752 is a nearby GC that has been extensively studied in the literature (see
Carretta et al. 2007 for references). 
It is the first GC where the O-Na anticorrelation has been discovered among
turnoff stars (Gratton et al. 2001).
The published values for its metallicity [Fe/H] range between $\sim$ -1.4 and
-1.6 (e.g., Pritzl et al. 2005). Here we adopt [Fe/H]~=~-1.5. 

In this work, we consider two sets of abundances for the stars of NGC~6752. 
The first one concerns the abundances of C, N, O and Na, measured 
by Carretta et al.~(2005)\footnote{Carretta et al.(2007) present the largest
sample of O and Na abundance determinations to date in NGC~6752. For seek of 
homogeneity we chose however to use the smaller sample of 2005 for which 
the abundances of C and N were simultaneously determined. This choice does not 
affect the present discussion.} 
on high resolution (R$\sim$40 000) UVES/VLT spectra; 
we keep only the measurements of the 8 dwarf stars of the sample, assuming it 
highly unlikely that they are contaminated by nuclearly processed material from 
their inner layers (Specifically the surface abundances of C and N are modified 
on the early RGB due to the first dredge-up of internaly CN-processed matter; 
e.g., Charbonnel 1994, Charbonnel et al. 1998, Gratton et al. 2000). 
The second sample consists of abundances of O, Na, Al, Mg, as well as Mg isotopes 
(\mga, \mgb, \mgc), measured on very high resolution (R$\sim$110 000) UVES/VLT 
spectra by Yong et al.~(2003), in 20 bright giant stars (Note that the 
surface abundances of these heavier elements are not modified during the 
first dredge-up event nor later on the RGB; e.g., Palacios et al. 2006).

The adopted abundances for  NGC~6752 are displayed in Fig.~1, as a function 
of the O abundance. 
Na, Al and \mgc \ clearly increase with decreasing O values, Mg and \mga \ decrease 
(slightly), while \mgb \ appears to be insensitive to O variations 
(as already discussed in Yong et al. 2003). 
We note that the variation of oxygen abundance is  larger in the case of 
Carretta et al.~(2005, a factor of $\sim$8)
than in the case of Yong et al.~(2003, a factor of 5). 
In the forme case indeed the authors have tried and enhance the chance of
finding extreme cases by selecting stars with likely strong and weak CN bands 
using the Str\"omgren $c_1$index.

\begin{figure*}
\centering
\includegraphics[angle=-90,width=\textwidth]{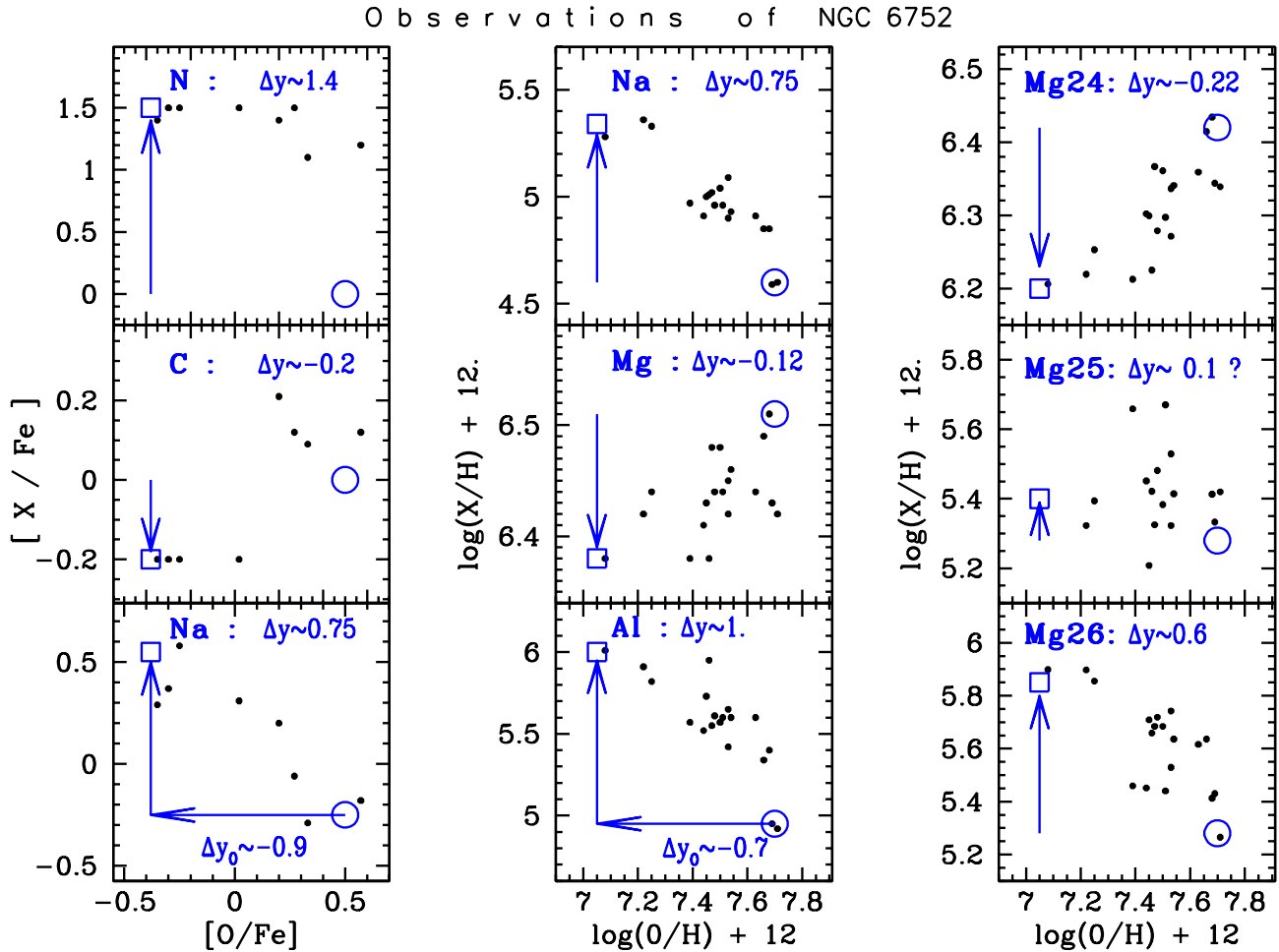}
\caption{Abundances of C, N, Na, Mg, Al and Mg isotopes in stars of NGC~6752, 
as a function of the corresponding oxygen abundance. Observations (filled symbols) 
in the {\it left column} (plotted as a function of [O/Fe]) are from 
Carretta et al.~(2005), and in the other two columns (plotted as a function of O/H) 
from Yong et al.~(2003). In all figures, the estimated initial abundances of the gas 
from the which  the cluster  was formed are indicated by an open  circle 
(i.e., it is assumed that they have the same composition as field halo stars of 
the same metallicity [Fe/H]=-1.5, with e.g. [O/Fe]=0.5, [C/Fe]=0, etc.). 
The most extreme abundances observed at present in NGC~6752 are indicated by open 
squares. Arrows indicate the magnitude and direction of the abundance spread, 
which is also given by $\Delta y$ (in dex, where $y$=log(abundance)) in each panel; 
$\Delta y_O$ in the bottom panels indicates the corresponding variation of oxygen 
abundance, larger in the case of Carretta et al.~(2005, a factor of $\sim$8), than 
in the case of Yong et al.~(2003, a factor of 5). 
\label{Fig1}
} 
\end{figure*}

The largest O values are similar to the ones of field halo stars of the 
same metallicity as NGC~6752;  we assume that this is the original O abundance 
of the intracluster gas.
This is also true for the corresponding values of Na, Al and Mg. We consider
then that all abundances corresponding to the highest O abundances (as indicated 
by open circles in Fig.~1) are the initial ones and that abundances at the other 
extreme (indicated by open squares) correspond to the most extremely processed 
material. For the initial abundances of C and N, we adopt [C/N] and [N/Fe]=0, 
in view of the corresponding abundances of field halo stars (Spite et al. 2005). 
We note that our adopted initial abundances for Mg isotopes, based on observations 
of NGC~6752 (Fig. 1, right panels) do not correspond exactly to the scarce available 
observations of field halo stars of the same metallicity. 
This ``mismatch'' has been noted by Yong et al. (2003, 2006) and is attributed 
by these authors to the action of a specific class of ``Mg-only polluters'' 
(assumed to be AGB stars), beyond the polluters required to explain the other 
composition anomalies. We do not consider here that idea , which introduces 
another level of complexity to a subject which is quite complex already. 
We simply note that (i) only in few  field halo stars the Mg isotopic ratios 
have been measured, and further observations are required before concluding on this 
(very important) topic; (ii)
Our adopted initial isotopic ratios are Mg24:25:26 = 87 : 6.5 : 6.5 
(see Fig. 1 and Table 1) i.e. they differ very little from
the ones  measured in NGC6752 by Yong et al. (2003), who give
83:10:7 for the highest ratio (corresponding to normal/unprocessed stars). 
In any case, those low values do not affect
the calculated final abundances of those isotopes or of $^{27}$Al, which are 
produced mostly from initial $^{24}$Mg in those temperatures.

For clarity, we display in Table 1 the assumed original and most extreme values 
for O, Na, Al, Mg, \mga, \mgb \ and \mgc, both by mass fraction and by number 
(in the log($N_H$)=12. scale), as well as the corresponding Mg24:25:26 ratios.

Our aim is then to find whether, starting with the original abundances $X_O$ 
(open cicles in Fig.~1), one may reproduce the observed extreme abundances $X_E$ 
(open squares in Fig. 1) for some, hopefully narrow, range of temperature T and 
timescale $\tau$ of H-burning. 

\begin{table}
\caption{Observed extreme abundances in NGC~6752}
\label{tabl1}
\begin{tabular}{rccccrr}
\hline
\hline
 & A$_{O}$ & X$_{O}$  & A$_{E}$ & X$_{E}$ & $\Delta$A & X$_{E}$/X$_{O}$ \\
\hline
O    &  7.65 & 5.4 10$^{-4}$   & 7.05  & 1.3 10$^{-4}$  & -0.60   &  0.25 \\
Na   &  4.60 & 6.9 10$^{-7}$   & 5.35  & 3.9 10$^{-6}$  &  0.75   &  5.6 \\
Al   &  5.00 & 2.0 10$^{-6}$   & 6.00  & 2.0 10$^{-5}$  &  1.00   &  10. \\
Mg   &  6.50 & 5.4 10$^{-5}$   & 6.38  & 4.7 10$^{-5}$  & -0.12   &  0.87 \\
\mga &  6.42 & 4.7 10$^{-5}$   & 6.20  & 2.8 10$^{-5}$  & -0.22   &  0.60 \\
\mgb &  5.25 & 3.4 10$^{-6}$   & 5.40  & 5.3 10$^{-6}$  & +0.15   &  1.5  \\
\mgc &  5.25 & 3.4 10$^{-6}$   & 5.85  & 1.4 10$^{-5}$  & +0.60   &  4.10 \\
\hline
\end{tabular}
\begin{tabular}{l}
Mg24:25:26 \ \ = \ \ 87 : 6.5 : 6.5 ($O$)   and 60 : 11 : 29 ($E$) \\
\hline
\end{tabular}
A=log(N/H)+12 ; $O$: Original ; $E$: Extreme
\end{table}

\section{Nucleosynthesis}

\subsection{The set-up}

\subsubsection{Nuclear reaction network}

We perform H-burning nucleosynthesis calculations at constant temperature T. 
Since all the reactions involved are 2-body reactions, density plays no role 
in the outcome, it only  affects directly the various timescales. For that reason, 
we display our results as a function of the consumed H mass fraction (the initial 
one being X$_H$=0.75) and not of time. We perform calculations for temperatures 
in the range T=25 to 80 MK (millions of degrees) that encompass typical burning 
conditions in a large variety of realistic stellar models (see \S 5).

The adopted reaction network extends from H to the Ca region and contains 
more than 140 nuclei, including very proton-rich ones. For instance, the range 
of O isotopes extends from $^{13}$O to $^{18}$O and for the Mg isotopes 
from $^{20}$Mg to $^{26}$Mg. It contains all the reactions of the cold and 
hot pp chains, the cold and hot CNO cycles, and all relevant interactions 
involving nuclei in the NeNa and MgAl mass ranges. The nuclide $^{26}$Al 
is treated as two distinct species:  the ground state $^{26}$Al$^g$ and the 
isomeric state $^{26}$Al$^m$, since those states are not thermalized at the 
relatively low temperatures of hydrostatic H-burning studied here. 
The communication between those states via gamma-ray transitions involving 
higher-lying $^{26}$Al levels is explicitly taken into account in our network. 
The required gamma-ray transition probabilities are adopted from 
Runkle, Champagne \& Engel~(2001).

Our reaction rate library is based on the evaluations of Angulo et al.~(1999) 
and Iliadis et al.~(2001) for the mass ranges of A$<$20 and A$\geq$20, respectively, 
with a few exceptions. For the $^{14}$N(p,$\gamma$)$^{15}$O reaction we use the 
recent rate of Runkle et al.~(2005) which is significantly lower than the rate 
given in Angulo et al.~(1999). New experimental information is also available 
for $^{17}$O(p,$\gamma$)$^{18}$F and $^{17}$O(p,$\alpha$)$^{14}$N. Our rates 
for these reactions are based on Fox et al.~(2005) and Chafa et al.~(2005), 
respectively. Finally, the $^{23}$Na(p,$\gamma$)$^{24}$Mg and 
$^{23}$Na(p,$\alpha$)$^{20}$Ne reaction rates are adopted from Rowland et al.~(2004). 
Note that the recommended rates for some of these reactions deviate strongly
from the NACRE evaluation (Angulo et al. 1999). For example, at 80~MK the new rates for 
$^{22}$Ne(p,$\gamma)^{23}$Na and $^{23}$Na(p,$\gamma)^{24}$Mg are factors 
of 103 and 33, respectively, smaller than the NACRE rates, while 
the new $^{27}$Al(p,$\alpha)^{24}$Mg rate is a factor of 10 larger than reported by NACRE. 

\subsubsection{Initial composition}
The initial composition of the mixture is the one of field halo stars of the 
same metallicity, [Fe/H]=--1.5, as NGC~6752 (see \S~2 and Table~1).
The abundances of the light isotopes are those resulting from Big Bang nucleosynthesis, 
with a $^7$Li abundance of log(Li/H)+12=2.65 (after BBM+WMAP; Steigman 2006).
The abundances of ONaMgAl are those suggested by the most O-rich stars in NGC~6752, 
namely those having [O/Fe]$\sim$0.5 (open circles in Fig.~1), while for C and N 
we adopt [X/Fe]=0. For the initial isotopic abundances, we adopt the values of Fig.~1 
(open circles) for the Mg isotopes (see also Table 1); observational data are 
not available for other isotopes, and we rely on Galactic chemical evolution 
calculations (Goswami \& Prantzos 2000) in order to evaluate the corresponding 
abundances at [Fe/H]=--1.5.

\begin{figure*}
\centering
\includegraphics[angle=-90,width=\textwidth]{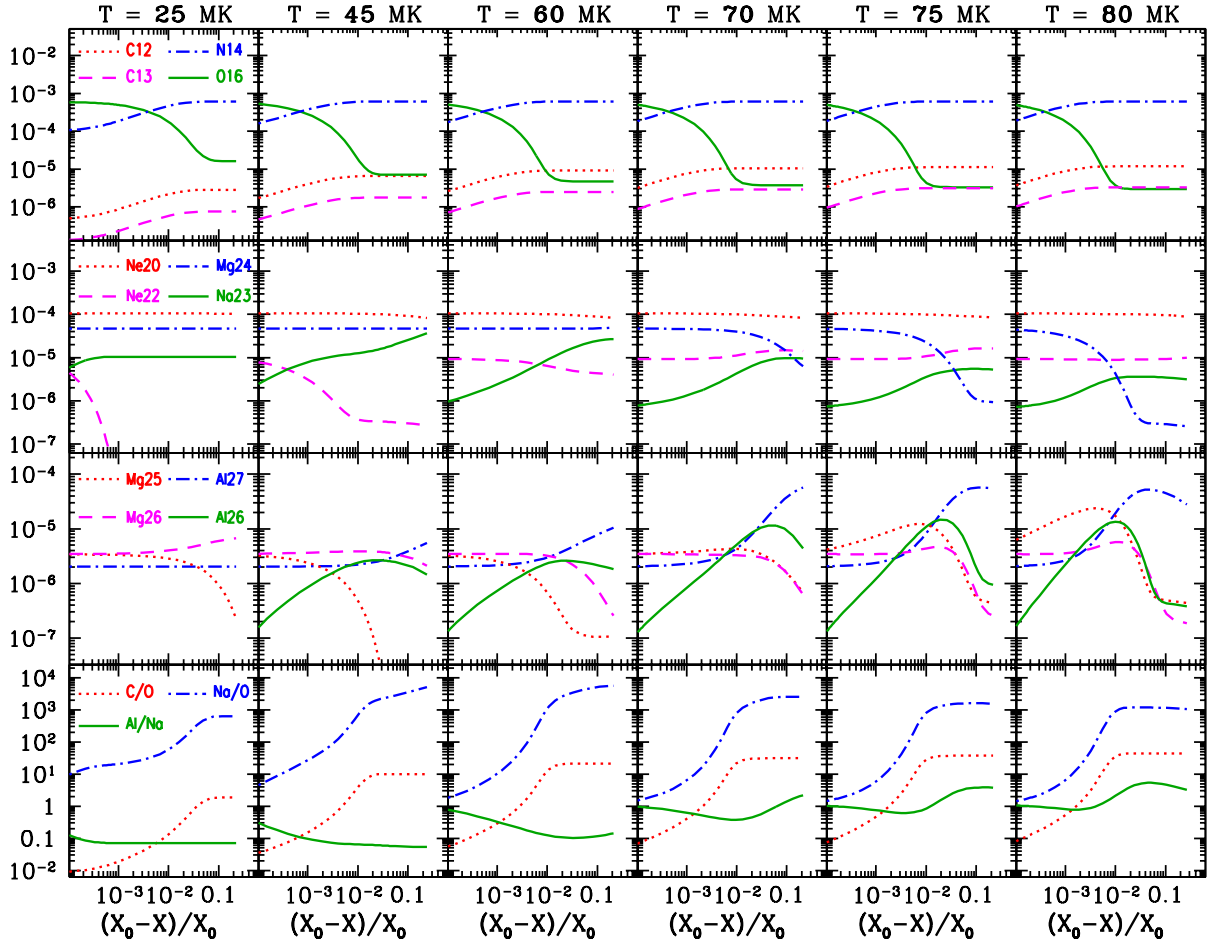}
\caption{Evolution of the composition during H-burning at constant temperature. 
Adopted temperatures are given on top of each column. Results are plotted 
as a function of the consumed H  fraction $\Delta$X/X$_0$ (where X$_0$=0.75), i.e.
time increases to the right of each panel. Abundances are mass fractions in the first 
three rows, while in the bottom row the quantity $(A/B)/(A/B)_{0}$ is given 
(where $A$ and $B$ are abundances by number and $0$ stands for abundances in the 
original mixture). 
\label{Fig2}
}
\end{figure*}

\subsection{Evolution of the composition during H-burning at constant temperature}

Results of our calculations for six different values of the temperature appear 
in Fig.~2. The abundances of selected CNONeNaMgAl isotopes
are plotted as a function of consumed mass fraction of hydrogen (X$_0$-X)/X$_0$. 
In the bottom row of that figure, the abundance ratios (by number) of C/O, Al/Na 
and Na/O are also displayed.

In the top row it is seen that $^{16}$O reaches its equilibrium value at T=25~MK after 
a few per cent of initial H is consumed, and at higher T after $\sim$1 \% of H 
is consumed; that equilibrium value is slightly temperature dependent. 
Isotopes involved in the CN cycle reach, of course, their equilibrium values 
much earlier.

In the second row, it is seen that $^{23}$Na is produced through the destruction 
of $^{22}$Ne already at 25 MK.
As temperature increases, a  fraction of $^{20}$Ne (less than 20\%) is converted 
to $^{22}$Ne and $^{23}$Na, which reach equilibrium above T$\sim$50 MK. 
The equilibrium value of $^{23}$Na is highest around T=50 MK and decreases slowly 
at higher T; at high temperatures, the equilibrium value of $^{22}$Ne remains 
$\sim$constant$\sim$0.1 of $^{20}$Ne. 

$^{24}$Mg is affected only above 70 MK, and only 1\% of its initial abundance is left at T=80 MK. At early times (when less than a few \% of H is consumed), its destruction leads to an increase of the abundances of \mgb \ and $^{26}$Al. When more than 10\% of H is consumed, almost all initial \mga \ has turned into $^{27}$Al, the abundance of which increases by more than a factor of 20. 

The two heavier Mg isotopes are more fragile than \mga \ and they are destroyed 
at low temperatures. When \mga \ is affected (above 70 MK) \mgb \ increases 
slightly early on, while \mgc \ remains $\sim$constant; at late times, they are 
both destroyed, following \mga.  Between the two epochs, substantial amounts 
of $^{26}$Al are produced. None of the Mg and Al isotopes reaches equilibrium, 
even when most of H is consumed.

The production of $^{26}$Al may be of importance for  \mgc, which is never produced 
as such (at least with the adopted set of reaction rates). If nuclear reactions 
cease (because material is brought to lower temperatures, through e.g. convection 
and stellar winds), then $^{26}$Al will decay to \mgc \ within 1 Myr, and the abundance 
of the latter will be found substantially enhanced. Note, however, that 
the \mgc/\mga \ ratio increases anyway, just because \mga \ is always destroyed.

The corresponding evolution of abundance ratios (by number) of C/O, Na/O and Al/Na 
is displayed in the bottom row of Fig.~2. It is seen that at high temperatures 
(above 60 MK) C/O and Na/O reach equilibrium values after a few \% of H is consumed, 
while enhanced Al/Na values are obtained only above T=70 MK.

In summary, our calculations show clearly that in the temperature range 70-80 MK 
the extreme abundances observed in NGC~6752 can be  qualitatively reproduced during 
the early  H-burning, i.e. after a few \% of H is consumed. O is strongly depleted 
always.  \mga  \ is also depleted, but  to a smaller extent. Na, Al and \mgc \ are 
largely enhanced (the latter in the form of $^{26}$Al). \mgb \ is very little 
affected in  that part of H-burning, but decreases considerably as more H is consumed. 
Finally, the sum of Mg isotopes (\mga+\mgb+\mgc+$^{26}$Al) is slightly smaller than the  
initial Mg amount, since some leakage takes place out of the MgAl chain.

We would like to emphasize that in this work we only consider recommended reaction
rates. The current rate uncertainties vary strongly from reaction to reaction. 
For example, at 80~MK the $^{24}$Mg(p,$\gamma)^{25}$Al reaction has a rate error 
of only 16$\%$ which makes it the best known rate among all proton-induced 
reactions on targets in the mass A$\geq$20 range (Powell et al. 1999). 
On the other hand, the rate errors for the reactions $^{22}$Ne(p,$\gamma)^{23}$Na,
$^{23}$Na(p,$\gamma)^{24}$Mg, $^{26}$Mg(p,$\gamma)^{27}$Al and $^{26}$Al(p,$\gamma)^{27}$Si 
amount to more than an order of magnitude at 80~MK. 

Izzard et al. (2007) present a list of such key reactions and  explore the effect
of the corresponding uncertainties on the outcome of AGB nucleosynthesis.
We plan to perform a similar exploration in the framework of our simple
one-zone parametrized models in a future work. Here, we simply note that
for the key reaction $^{26}$Al(p,$\gamma$)$^{27}$Si (the latter decaying 
quasi-instantaneously to $^{27}$Al),  Izzard et al. (2007) suggest uncertainties by 
factors f= 0.5 to 600 over the value of Iliadis et al. (2001) adopted here.
By adopting f=0.5 we find that higher temperatures than 80 MK are required to
obtain the extreme $^{27}$Al values, while for f=50 temperatures lower than 70 MK
are required; in the latter case, however, we never get enough \mgc ~(since
there is never enough $^{26}$Al). Such temperatures are outside the
temperature range that we determine in the next section. In view of that, one may  
conclude either that:

(i) the true reaction rate values are close to those adopted here, which lead 
to results agreeing with observations for the temperature range T$\sim$74 MK 
(see next section), or 

(ii) some other combination(s) of reaction rates may also satisfy observations,
perhaps for some different temperature range.

We keep option (i) here, and 
we explore the impact of reaction rate uncertainties on our models in future work. 

\subsection{Constraining the temperature range}

After the first (and broadly encouraging) estimates based on the results of Fig.~2, 
we attempt  to constrain more quantitatively our parameter space, using the observables 
of NGC~6752. From the results displayed in  Figs.~1 and 2, we note that:

a) Oxygen is the key element, both for observational reasons (it is extensively observed) 
and for theoretical ones (its abundance varies widely). The regularity of the observed 
abundance patterns (abundances vary monotonically with O and with relatively little scatter) 
suggests that the O abundance in the processed matter has reached its equilibrium value 
(otherwise, considerable dispersion in abundance patterns would be expected) ; 
consequently, {\it equilibrium values of O should be adopted from Fig.~2}.

b) The equilibrium values of O at high T are $\sim$100 times lower than 
the initial ones; but the observed lowest O abundances in NGC~6752 are 
only $\sim$5-8 times lower than the original value (depending on whether 
the Carretta et al~(2005) or the Yong et al.~(2003) data are used). 
This means that even the  extreme observed abundances result necessarily 
from a {\it mixture of processed material with original one}; in other words 
material coming directly from the H-burning regions is never observed directly, 
otherwise O abundances should be much lower than actually observed. 
Obviously, intermediate abundance values between $X_O$ and $X_E$ should result 
from various degrees of such mixing.

\begin{figure}
\centering
\includegraphics[width=0.49\textwidth]{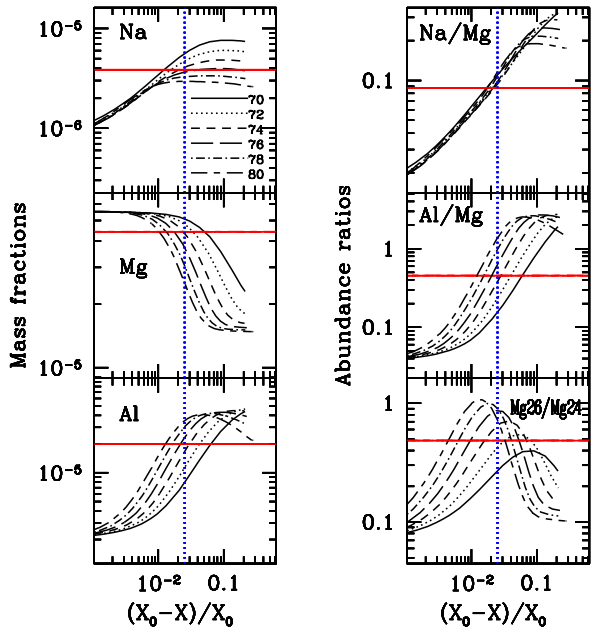}
\caption{Evolution of mass fractions (left column) and abundance ratios by number 
(right column) of elements in the Na-Al region, during H-burning at constant 
temperature; they are plotted as a function of the consumed H fraction (see Fig.~2) 
for six values of the temperature in the 70-80 MK range (see {\it top left panel}). 
The abundance of $^{26}$Al is always counted in the one of \mgc \ and of total Mg. 
All theoretical values are corrected for mixing with original material with a factor 
$f$=0.31, such as to ensure that the  abundance in the mixture will be equal to 
$X_{E,OBS}$ for oxygen (see text).
{\it Horizontal solid lines} in all panels represent the observed extreme values of each 
element or abundance ratio ($X_E$ in Table 1). The {\it vertical dotted lines} in all 
panels indicate a consumed H fraction of $\Delta X/X$=0.025, where calculations 
at T=74-76 MK agree with observational requirements for all elements. 
\label{Fig3}
} 
\end{figure}

c) From the previous (mixing) argument, one concludes that no direct comparison is possible between
observations  (the most extreme values, to the left of all panels in Fig.~1, or X$_E$ in Table 1) and the theoretical results displayed in Fig.~2. However, by combining theory (i.e. the relatively stable - at high T - equilibrium values of O abundance $X_{EQU}$) and observations (the observed extreme values of O $X_{OBS}$) one may derive the {\it required mixing factor } $f$; indeed, the observed values being a mixture of one part of processed material with $f$ parts of original one, one has
\begin{equation}
X_{OBS} \ = \ {{X_{PROC} \ + \ f \ X_{ORIG}}\over{1 \ + \ f}}
\end{equation}
from which 
\begin{equation}
f \ = \ {{X_{OBS} - X_{PROC}}\over{X_{ORIG} - X_{OBS}}}
\end{equation}
In the case of oxygen one has:  for the original value $X_{ORIG}$=5.4 10$^{-4}$ 
(Table 1), for the  extreme observed value of O $X_{E,OBS}$=1.3 10$^{-4}$ 
(Table 1, for the Yong et al. 2003 data) and  for the equilibrium value at T=70-80 MK 
$X_{PROC}$=3.5 10$^{-6}$ (from Fig.~2).
From those values one obtains $f_O$=0.31 for oxygen. Applying that mixing (=dilution) 
factor to all other theoretically obtained  abundances through Eq. 1, one may derive 
then the corresponding {\it  abundances in the mixed material} $X_{E, OBS}$ for all 
other elements of the network. The question then is: are those mixed abundances 
(which, by construction, reproduce the extreme observed oxygen abundance) compatible 
with the extreme observed values of all other elements and isotopes appearing in Fig.~1? 
For what temperatures and consumed fractions of H ?
Is there a range of T and $\Delta X$ satisfying all the observational constraints?

The resulting abundances in the mixture (i.e. the theoretical ones, corrected for 
mixing with $f$=0.31) appear in the various panels of Fig.~3 in a way similar 
to those displayed in Fig.~2, but this time a more refined grid of temperatures 
is used, with steps of $\Delta$T=2 MK in the temperature range 70 to 80 MK. 
The purpose of that exercice is to determine more precisely the ranges of temperature 
and consumed H fraction that satisfy simultaneously all observational requirements; 
the latter, i.e. the values X$_{E,OBS}$ appear as horizontal solid lines in Fig.~3.  
In other terms, we seek to find whether  {\it all} the observed extreme abundances 
in NGC~6752 can be explained in terms of nucleosynthesis occuring in a narrow range 
of (constant) temperatures, after some mixing with original material.
However, since abundances in the Mg-Al region do not reach equilibrium but evolve 
steadily with time (or H mass fraction), we have to seek for specific solutions 
in the T vs $\Delta X$ plane; if all abundances were in equilibrium, the situation 
would be much simpler and only a range of temperatures would be sought. 

An inspection of Fig.~3 shows that Na is not a good ``thermometer'', since its 
quasi-equilibrium abundance varies little (just a factor of 3) between 70 and 80 MK. 
\mga is much more sensitive to temperature: after a  few \% of H is consumed, 
\mga / abundance  decreases rapidly, especially for T$>$74 MK. 
The opposite is true for $^{27}$Al and for \mgc \ (which does not appear in the figure): 
their abundances increase rapidly with $\Delta X$ (at least for $\Delta X <$0.1) 
and with T.

These trends, namely the stability of Na with $\Delta X$, the decrease  of Mg 
(mainly the form of \mga) and the increase of  Al and \mgc, allow one to constrain 
the range of both T and $\Delta X$.
An inspection of Fig.~3 shows that towards the middle of the explored temperature range, 
between 74 and 76 MK, and for consumed H fraction $\Delta X$/X$\sim$2-3 \%, 
the results of the calculations for the mixture are quite close to the observed 
extreme values, for all the involved elements. Thus, at least in the framework 
of our idealised study (constant T), there is a formal solution to the problem 
we tackled. Taking into account the various uncertainties of that problem 
(nuclear reaction rates at such low temperatures, observational errors) and 
the fact that temperatures do not remain constant in realistic nucleosynthesis sites, 
we consider this formal solution as extremely satisfactory. We note that the constraints 
we derived on temperature are rather strong, in view of the strong dependence of nuclear 
reaction rates on that parameter; however, the physical significance of our derived
temperature in the case of convective zones will be analysed in Sec. 5.1. 
On the contrary, the constrains on $\Delta X$ are obviously unrealistic, since in a 
astrophysical site, some mixing of material from the H-burning zone with fresh H 
(and other metals) from external zones is always expected; $\Delta X$ can be hardly 
defined then.

\begin{figure*}
\centering
\includegraphics[angle=-90,width=\textwidth]{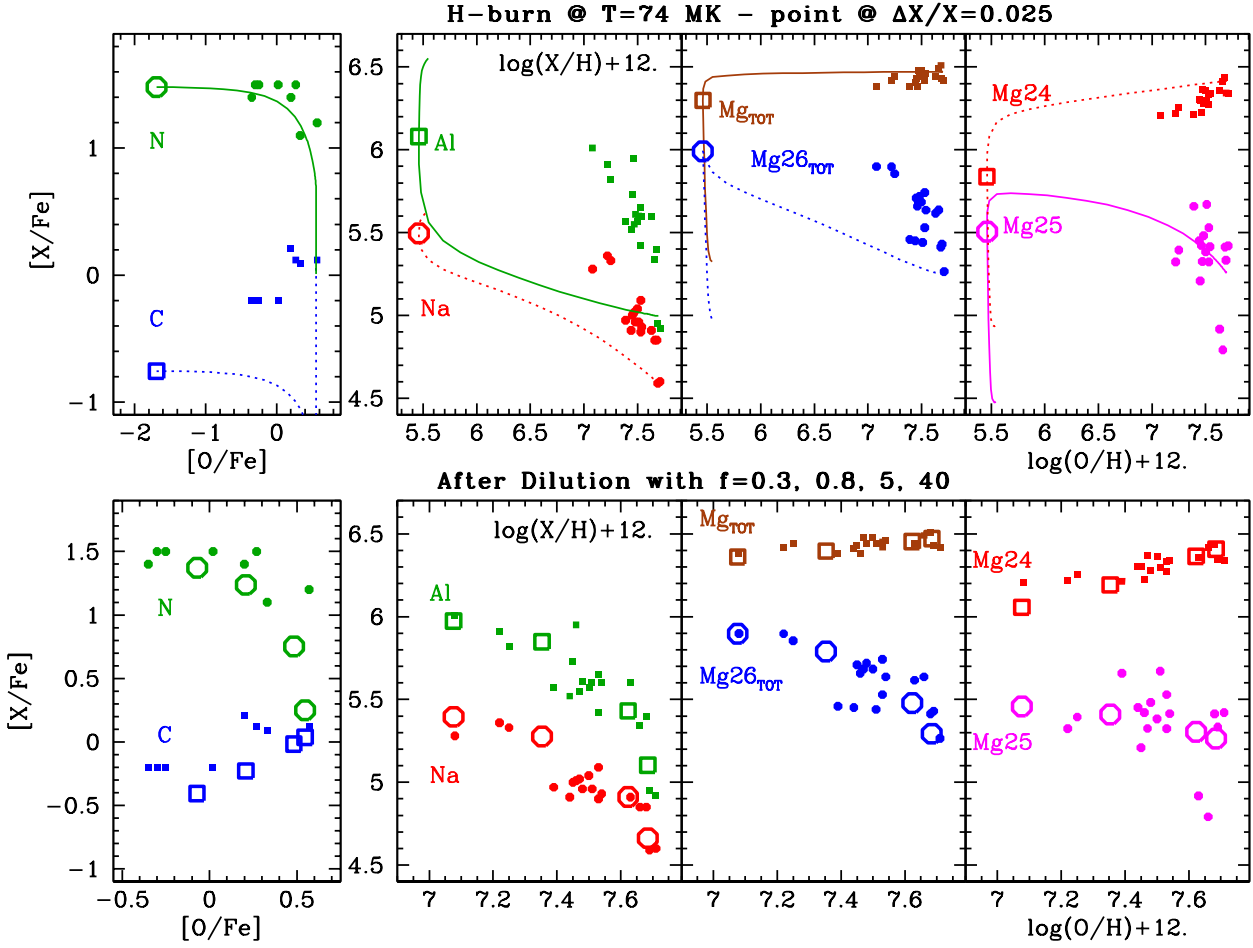}
\caption{Abundances of C, N, Na, Mg, Al and Mg isotopes as a function of O abundance. 
The O abundance scale is expanded in the upper row, to accomodate the range of values (i.e. down to the O equilibrium value) obtained
in the calculation at T=const=74 MK; results of calculations are shown by the solid and dotted curves in each panel,
with O reaching its equilibrium value in the extreme left. Abundances corresponding to a consumed H fraction of
$\Delta$X/X$_0$=0.025 are shown with open circles and squares in each panel in the upper row.
Observations (filled points in all panels) are from references given in Fig.~1. Open symbols in the lower row
panels denote abundances obtained after mixing the calculated abundances at $\Delta$X/X$_0$=0.025 of the corresponding upper  
panels with various amounts of
material having the original composition. The corresponding dilution factors appear on top of the lower row panels; smaller dilution factors result in abundances closer to the processed ones (e.g. higher Na, Al etc.). The abundances of Mg$_{TOT}$ and \mgc$_{TOT}$ include those of 
$^{26}$Al.
\label{Fig4}
} 
\end{figure*}

\subsection{Mixing of processed with original material}

Once the extreme observed abundances $X_{E,OBS}$ are understood (as a result of mixture 
of one part of material processed at T$\sim$75 MK and $\Delta X \sim$0.025 with 
$\sim$0.30 parts of original one), all other observed abundances can be recovered by 
assuming mixing with various dilution factors (larger than 0.3). 
The results of that exercice appear in Fig.~4. 

In the upper panel of Fig.~4 we first display again the results of calculations 
at T=74 MK, but this time the curves are plotted as a function of the O abundance; 
it can be seen that the results of the calculations  deviate a lot from the 
observations (points), as O decreases rapidly towards its equilibrium value.

Selecting the point corresponding to $\Delta X$/X$\sim$0.025 (in the left of all 
panels of Fig.~4,  after the analysis of \S~3.2)  and mixing that composition 
with original material to various degrees (dilution factors f=0.3, 0.8, 5. and 40, 
respectively) we obtain the series of open symbols in the bottom panels of Fig.~4. 
Values to the left correspond to f=0.3, i.e. 1/3 of original material for every part 
of processed one. An element heavily depleted in the H-processed zone, like oxygen, 
will have $\sim$1/4 of its original abundance in the mixture; on the other hand, 
elements largely overproduced in the H-burning zone, like Na and Al, will have their 
processed abundances reduced by $\sim$30\% in that same mixture. Those values 
(with $f$=0.3) correspond to $X_{E,OBS}$, as discussed in \S~3.3; for larger values 
of $f$, the composition gets closer and closer to the original one. Overall, 
the obtained mixtures reproduce quite satisfactorily the full range of observations. 

A comparison of the upper and lower panels in Fig.~4 shows clearly the important role of 
dilution in getting the observed abundances from those in the H-burning zone: the latter 
(caracterized by extremely low equilibrium values of oxygen) are not at all reminiscent 
of the observed abundance values, even the most extreme ones. Only when mixing of 
processed and original material is assumed, the full range of observed values is 
recovered. In the case of NGC~6752 considered here, the  abundances  of each and 
every one of the observed CNONaMgAl elements and Mg isotopes are reproduced extremely 
well, for a very restricted range of H-burning temperatures and fraction of H consumed. 
It is not clear whether such a highly idealized situation can occur in a realistic site; 
we discuss that point in \S~5.

\begin{figure}
\centering
\includegraphics[width=0.5\textwidth]{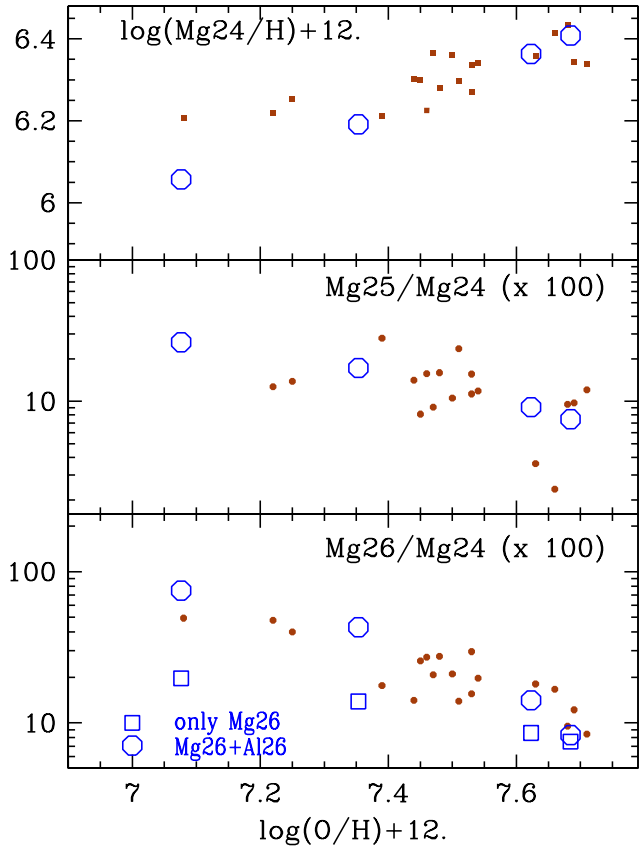}
\caption{Mg isotopic ratios vs O/H. Theoretical results (open symbols) are
obtained for T=74 MK and at $\Delta X$/X=0.025, as in Fig.~4 ; they are mixed 
to various degrees with material of original composition (mixing factors $f$=0.3, 0.8, 
5 and 40, from left to right). Observations (filled points) are as in Fig.~1.
\label{Fig5}
} 
\end{figure}

In Fig.~5 we take a closer look to the situation concerning the Mg isotopes, 
since they play a critical role in the determination of the temperature range 
(due to the decrease of \mga \ at high T and $\Delta X$). Indeed, Al and Na behave 
rather similarly (they both increase with T and $\Delta X$) and are not sufficiently 
sensitive to pinpoint T.

In Fig.~5 it is seen that, in the selected conditions of T and $\Delta X$,  \mga \ 
is slightly underproduced w.r.t. observations (by 0.15 dex), while the \mgc/\mga \ 
ratio is slightly overproduced. We note that  $^{26}$Al is counted in the abundance 
of \mgc (and of total Mg, as in Fig.~4) and is at the origin of the large \mgc/\mga \ 
ratio; if $^{26}$Al is neglected, then the extreme abundances of \mgc/\mga \ 
in NGC~6752 are underpredicted by a factor of 2.5 ({\it open squares} in Fig.~5).

\section{Lithium and Fluorine in globular clusters}

In the previous section, mixing of processed with {\it original} material was invoked, 
in order to explain why the observed oxygen abundances (even the most extreme ones) 
are far above the processed  equilibrium value. In actual reality, this constraint 
does not require completely unprocessed  (i.e. truly pristine) material, only  
{\it material  processed below} $\sim$20 MK, i.e. with unaltered abundances of O 
and heavier elements. Since a large fraction of the mass of a star is found 
at temperatures lower than 20 MK, our analysis so far does not help to clarify 
the nature of the unprocessed material: is it pure interstellar medium, 
or can it be just part of the stellar envelope ? 


A crucial constraint, recently discovered, can help to distinguish between the two 
alternatives. Pasquini et al.~(2005) reported observations of Li in 9 turn-off stars 
of  NGC~6752 with UVES (see Fig.~6) and found that Li is correlated with O and anticorrelated 
with Na. Stars with ``original" O (high) and Na (low) have a Li content similar 
to field halo stars in the Spite plateau, i.e.
$A$(Li)=log(Li/H)+12$\sim$2.3. Stars with ``extreme" abundances (lowest O and 
highest Na) {\it are not devoid of Li} but have  a significant amount of $A$(Li)$\sim$2. 
In view of the fragility of Li, which ``burns" at temperatures higher than $\sim$2.2~MK, 
this discovery implies that the ``unprocessed" material of our mixture has always been 
kept cooler than 2.2~MK. Since the corresponding stellar zones are a tiny fraction 
of the mass of any star, it appears natural to conclude that the ``unprocessed" 
material of our mixture is plain ISM.

In Fig.~6 we repeat the procedure followed in \S~3.3 and Fig.~4, this time for Li 
and Na. Material processed at T=74 MK (until a fraction of H $\Delta X$/X=0.025 
is consumed) is mixed with $f$ parts of original material, which has $A$(Li)=2.65 
(i.e., the cosmological value after WMAP + SBBN; Steigman 2006). 
By varying $f$ between 0.1 and 
some large value (above 50) we naturally obtain a decline of the Li abundance 
with increased Na abundance  (open circles in  Fig.~6). 
This corresponds to the Li abundance with which each individual GC star is born.
Assuming then that in each star Li has been uniformly depleted by 0.4 dex from 
its initial value determined previously, as is the case for halo stars in the 
Spite plateau (e.g., Charbonnel \& Primas 2005), we obtain the filled circles 
in Fig.~6. This corresponds to the predicted present Li abundance 
in the turnoff GC stars that can be directly compared with the observational data. 

\begin{figure}
\centering
\includegraphics[width=0.5\textwidth]{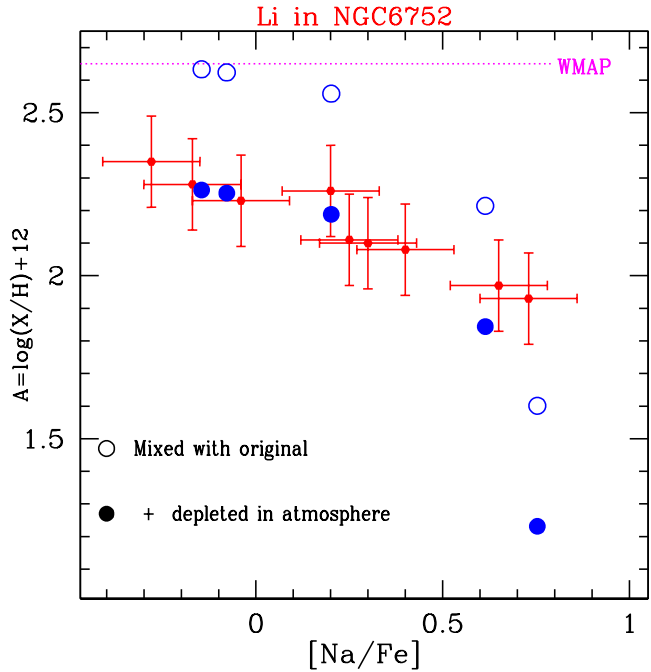}
\caption{Li in the globular cluster NGC~6752. Observations (filled symbols with 
error bars) are from Pasquini et al.~(2005) derived for the Alonso et al.~(1996) 
temperature scale. Open symbols indicate H-burned material 
($\Delta X\sim$0.025 of H consumed and no Li present), mixed to various degrees with 
material of original composition (containing Big Bang Li, at the level of WMAP); 
dilution factors are 0.1, 0.6, 5, 40 and 250, from left to right. Filled symbols 
indicate uniform depletion of 0.4 dex in all theoretical mixtures, to account 
for the difference between the WMAP constraint on primordial Li and observations 
of the ``Spite plateau" (Charbonnel \& Primas 2005).
\label{Fig5}
} 
\end{figure}

Pasquini et al.~(2005) already interpreted the presence of Li in the most Na-rich 
stars of NGC~6752 as an additional clue in favor of the self-enrichment scenario.
These authors suggest that Li must have been created by the same stars that
build up the other abundance anomalies, and discuss the possible contribution of AGB 
stars (see Ventura \& D'Antona 2005a). Our analysis convincingly shows however 
that Li production by the stellar 
polluters is not mandatory in the case of NGC~6752 where the abundance patterns are 
nicely reproduced by mixing to various degrees pristine ISM with H-processed 
(and totally Li-free) material.This supports the results of Decressin et al.~(2007)
who successfully explain the chemical inhomogeneities in NGC~6752 
through mixing of the wind of fast rotating massive main sequence stars with pristine 
material by using a dilution factor derived from the Li behaviour.
Last but not least the predictions shown in Fig.~6 provide additional support to 
the idea that old metal-poor low-mass stars manage to destroy part of their initial 
Li abundance in a very uniform way, both in the halo and in GCs 
(Talon \& Charbonnel 2004; Richard et al. 2005; Korn et al. 2006; Charbonnel 2006).

Pasquini et al.~(2005) underline the different Li abundance patterns found 
in NGC~6752 and NGC~6397 and which may be important to pinpoint the processes 
involved  in the self-enrichment scenario. In NGC~6397 indeed the turnoff stars observed 
to date all share the same Li abundance with essentially no intrinsic scatter 
(Pasquini \& Molaro 1996; Th\'evenin et al. 2001; Bonifacio et al. 2002; 
but see also Korn et al. 2006), although large variations in C,N,O, and Na were 
uncovered in this GC (Carretta et al. 2005). 
We note however that in NGC~6397 the stars for which both Li and Na could be determined 
cover only a modest range of the Li-Na anticorrelation seen in NGC~6752: 
they have [Na/Fe] between +0.09 and +0.35, and in this Na range Li is also
almost constant in NGC~6752 as can be seen in Fig.~6. 
Li determinations for stars exhibiting stronger Na enrichment are urgently needed 
in NGC~6397 as well as in other GCs before one can conclude on possible 
(and potentially existing) variations in the self-enrichment scenario 
from cluster to cluster.


\begin{figure}
\centering
\includegraphics[width=0.5\textwidth]{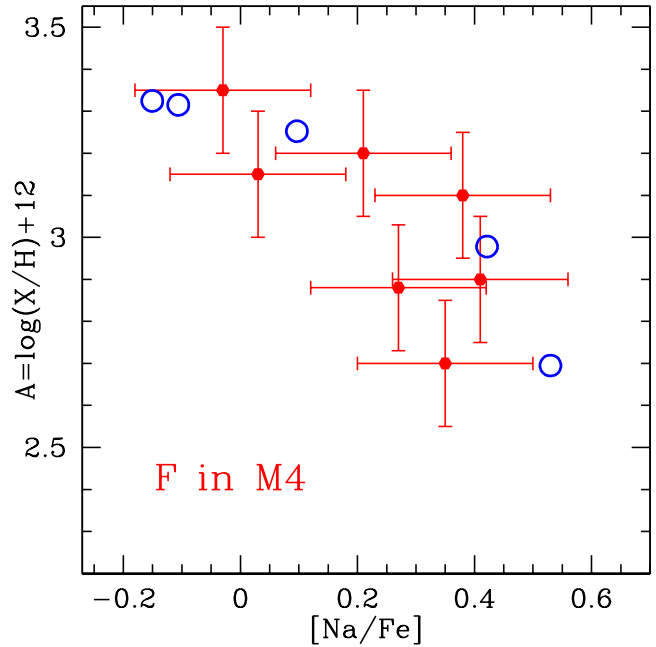}
\caption{Fluorine in the globular cluster  M4. Observations are from Smith et al.~(2005).
Open symbols indicate H-burned material at T=74 MK ($\Delta X\sim$0.025 of H consumed 
and no F present), mixed to various degrees with material of original composition; 
dilution factors are 0.1, 0.6, 5, 40 and 250, from left to right. 
\label{Fig6}
} 
\end{figure}

Fluorine, another fragile element (albeit less fragile than Li) has  been observed 
by Smith et al.~(2005) in seven red giants of M4, in high resolution infrared 
spectra obtained with the Gemini South telescope. Fluorine is usually destroyed 
in high temperature H-burning  and its abundance in M4 is again anticorrelated 
with the one of Na (Fig.~7): an increase by a factor of $\sim$3 in Na corresponds 
to a decrease of the F abundance by  factor of $\sim$4. 
Applying exactly the same prescription as in the case of Li, i.e. by diluting 
material processed at 74 MK (rich in Na and having no F) with $f$ parts of 
pristine material (poor in Na and assumed to have log(F/H)+12=3.35), we obtain 
the open symbols in Fig.~7. Again, the full range of observed values is 
(unsurprisingly) reproduced. 
Smith et al.~(2005) interpret the F-Na anticorrelation as the result of pollution 
of the intracluster gas by relatively massive AGB stars that destroy F in the 
course of their evolution. We note however that the observed fluorine abundance 
variations do not provide a tight constraint to the mass of the polluters and
does not help to discriminate between massive AGB stars and massive stars (see 
\S~5) which both destroy this element in their interior.

The conclusion of this section is that observations of fragile elements (Li, F) 
in GC stars are naturally understood in the framework of the ``mixture scenario'' 
explored here. This has already been mentioned in previous works (Pasquini et al. 
2005; Decressin et al. 2007) but our study is the first to show it 
quantitatively. The advantage of those observations (especially Li) is that they 
clearly indicate that the unprocessed material has to be of pristine nature 
(i.e. pure ISM) rather than from the stellar envelope of the polluter itself; 
this information cannot be obtained from the abundance patterns of the  heavier 
elements, analysed in \S~3. However, contrary to the case of the heavy elements, 
observations of fragile elements offer no clue as to the temperature range of 
the processed material and can not be used to pinpoint the mass of the stellar 
polluters. 

\section{Temperature and nucleosynthesis sites}
\subsection{H-burning temperature in stars}

The analysis of \S~3 is specifically adapted to the case of NGC~6752. 
In the framework of our extremely simplified model (nucleosynthesis at 
constant temperature, mixing with pristine material), that analysis allows us 
to determine a quite narrow range in the (T, $\Delta X/X$) plane where H-burning 
products reproduce the full set of observational data for NGC~6752, after 
appropriate mixing. 

The ``constraint'' on $\Delta X$ is rather meaningless, since in most realistic 
situations (convective cores of massive stars, bottoms of AGB convective envelopes, 
see below) hydrogen is constantly mixed in the H-burning zones, both replenishing 
(at least partially) the H-content of the zone and diluting the abundances of 
the H-burning products.
It is important, however, to find that the observational data are reproduced 
{\it for the same value of } $\Delta X$ (no matter what that value is), just for 
consistency reasons : if e.g. at a given T, the observed Na excess was reproduced 
only early on (at low $\Delta X$) and the observed Al excess only at late times 
(large $\Delta X$), no meaningful conclusions about T would be drawn. 

The constraint on temperature T is quite robust, since nuclear reaction rates 
display such a strong temperature dependence: for instance, for T$<$70 MK, \mga \  
is hardly affected at all, while for T$>$80 MK it is rapidly destroyed. 
Although H-burning temperatures do not remain absolutely constant during stellar 
evolution, they vary relatively little in quiescent burning stages. 
It is meaningful then to ask in which kind of realistic stellar environment(s) 
correspond the temperatures of 74-76 MK that we determined through the analysis 
of \S~3.

Maximum H-burning temperatures corresponding to various stages of stellar evolution 
are displayed in Fig.~8 as a function of initial stellar mass.
The theoretical values are given for models with [Fe/H]=--1.5. 
Massive stars (i.e., with an initial mass $\geq$ 20~M$_{\odot}$) are from 
Decressin et al.~(2007) while low- and intermediate-mass stars come from 
Decressin et al.~(in preparation) and Siess \& Pumo (2007); 
for massive AGBs we also show some predictions by Ventura \& D'Antona~(2005b).
In the case of core H-burning we show the range covered by the central temperature
during the whole main sequence (vertical straight lines). 
For the other phases we show only the maximum temperature reached in the H-burning
regions of interest : in the H-burning shell during the red giant phase (i.e., 
at the ``tip" of the RGB; open squares) and during the central He-burning phase 
(i.e, on the ``clump"; open circles), and at the bottom of the convective envelope 
in AGB or super-AGB stars (filled squares for the models by Decressin et al. 
and Siess, filled circles for the models by Ventura \& D'Antona).

\begin{figure}
\centering
\includegraphics[width=0.5\textwidth]{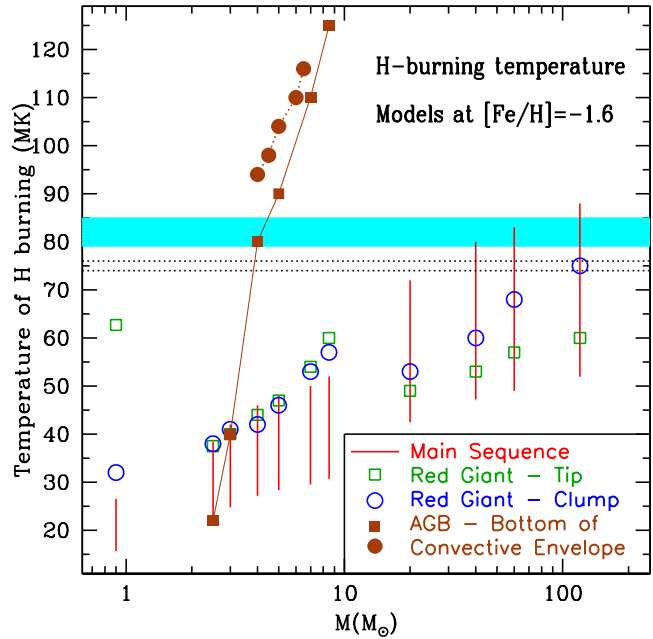}
\caption{H-burning temperatures at various phases of a star's life, as a function of 
initial stellar mass. The correspondence between symbols and evolutionary phases 
is displayed in the bottom right part of the figure. 
Vertical straight lines (for central H-burning) start at the ZAMS and end at central
H-exhaustion. For the other phases we show only the maximum temperature reached 
in the H-burning zones of interest (see details in the text). The two {\it horizontal 
dotted lines} enclose the temperature range of our one-zone models reproducing the extreme abundances 
that are observed in NGC~6752, after the analysis of \S~3. The {\it shaded
aerea} encloses our rough estimates for the corresponding maximal temperature range in the case of convective H-burning 
regions (in the bottom of AGB envelopes or in the center of massive star cores), as discussed in Sec. 5.1.
\label{Fig7}
} 
\end{figure}

The horizontal dotted lines enclose the temperature range reproducing the 
extreme abundances that are observed in NGC~6752, as discussed in \S~3. 
However, this temperature range is not necessarily directly comparable
to those of the stellar models.
Obviously, for nuclear burning in radiative (and narrow) regions, the temperature
of our calculations  is 
simply  the temperature of those regions. But for convective regions (such as 
the cores of massive stars or the Hot-Bottom  burning envelopes of AGBs) 
the situation is different : burning takes place
essentially in the hottest region (the massive star center  or the bottom of 
the  AGB envelope) but it involves fuel from all over the convective region
(because convective timescales are shorter than nuclear ones).
In those conditions it can be shown that the fuel burns with
an effective reaction rate
\begin{equation}
R_{eff} \ = \ {{1}\over{M}} \ \int R(T(m)) \ dm
\end{equation}
where $R(T(m))$ is the reaction rate at the shell of mass coordinate $m$ 
and temperature $T(m)$, $dm$ is the mass of that shell and $M$ 
the total mass of the convective region (see e.g. Eq. 5 in Prantzos et al. 1986).
In that integral, the largest contribution comes from the hottest (highest T)
region (because reaction rates are extremely sensitive to temperature and 
contributions from all cooler regions may be considered negligible), but
multiplication by $dm/M$ leads then to a smaller effective rate, corresponding to lower T.

Our calculated rates obviously simulate $R_{eff}$ of convective regions, and 
thus allow us to constrain only the temperature $T_{eff}$ which corresponds to 
$R_{eff}$. This is only a lower limit to the actual temperature $T_b$ where 
burning  takes place (i.e. the star center or the bottom of the AGB envelope).
How much larger may be that temperature $T_b$ ? It depends on the actual temperature
profile of the convective region. An approximate estimate can be made
as follows:

Assuming that  $R(T)\propto T^L$ (with e.g. $L$=20 for CNO cycle reactions)
and that  $ dm(T_b) = M/N$  (the hottest shell of temperature $T_b$ 
has a mass $dm$ which is  $N$ times smaller than the mass $M$ of the convective region)
one finds that :  $T_b/T_{eff} \ = \ N^{1/L}$
For instance, for $L$=20 and $N$=10, one has $T_b/T_{eff}$ \ = \ 10$^{0.05} \sim$1.12
Thus, our $T_{eff}$=74 MK corresponds to a more ``realistic'' peak temperature
of the convective region $T_b$=83 MK.
Since NeNaMgAl nuclei have higher Coulomb bariers than CNO nuclei, $L$ must be 
slightly larger than 20 and $T_b$ slightly smaller than derived above.

Our conclusion then is that the actual maximum temperature of 
convective H-burning zones which
leads to the observed extreme abundances in NGC6752 is $\sim$10-15\% higher than
the temperature determined here, i.e. in the 80-85 MK range. 
Note, however, that reaction rate uncertainties also impact on that 
derivation.

\subsection{Low-mass stars presently observed in GCs}

As already mentioned, the temperature inside the GC turnoff stars that 
we observe today (with a typical mass of the order of 0.8-0.9M$_{\odot}$, 
see the corresponding vertical line in Fig.~8) is too low for the proton-capture 
reactions to operate in the NeNa- and Mg-Al chains, implying that the observed 
abundance variations cannot be of intrinsic nature. While low-mass stars climb 
the RGB the temperature increases inside the H-burning shell surrounding the 
degenerate He-core;  it reaches relatively high values (of the order 
of $\sim$ 60MK for the 0.9M$_{\odot}$ 
model) that are sufficient for the NeNa-chain to operate and for the two heavier 
Mg isotopes to burn around the H-burning shell deep inside the star
(Kudryashov \& Tutukov 1988; Denisenkov \& Denisenkova 1989, 1990; 
Langer et al. 1993; Langer \& Hoffman 1995; Cavallo et al. 1999; Weiss et al. 2000). 
However the observed variations of $^{24}$Mg require still higher internal 
temperature (see also Arnould et al. 1999).
These difficulties, together with the finding that the GC chemical patterns 
(i.e., anticorrelations) are similar in turnoff and RGB stars, imply that the 
abundance anomalies cannot be generated in the course of the evolution of 
the long-lived objects that we are currently observing. In other words, 
the so-called {\sl {evolution hypothesis}} that calls for non-canonical 
very deep mixing inside the low-mass star itself should be discarded. 

\subsection{AGB stars}

It is usually claimed that hot bottom burning (hereafter HBB) in massive AGB stars 
undergoing thermal pulses is responsible for the observed composition anomalies in GCs. 
As can be seen in Fig.~8 indeed, the relatively high H-burning temperatures 
required to reproduce the extreme NGC~6752 abundances are encountered at the bottom
of the convective envelope of these objects.
The region in the envelope that is hot enough for H-burning is relatively thin; 
however thanks to the very efficient convective mixing the matter in the entire 
envelope goes through the burning region a large number of times ($\sim$ 1000) 
during each thermal pulse cycle. This directly affects the surface and wind 
composition of the stars.
Note that for a given initial stellar mass the temperature at the bottom of the 
convective envelope T$_{bce}$ increases from pulse to pulse on the AGB 
and reaches the maximum value shown in Fig.~8 after a few thermal 
pulses (see e.g., Ventura \& D'Antona 2005a). As the star evolves and loses mass
T$_{bce}$ decreases again.

The difference  in the theoretical predictions of the two sets of models considered 
here underlines   the uncertainties associated with the input physics that is adopted 
in the computations. Here it comes mainly from one of the most relevant uncertainties 
connected to stellar evolution, i.e., the treatment of convection that strongly 
influences the stellar characteristics during the TP-AGB phase : 
Decressin and Siess models use the traditional MLT (Mixing Length Theory) 
while Ventura \& D'Antona use the FST (Full Spectrum of Turbulence; 
Canuto et al. 1996).  In the latter case the convection 
efficiency is stronger, leading to higher temperatures at the bottom of the convective 
envelope of the AGBs (see Ventura \& D'Antona 2005a for a more detailed discussion). 
Other key (but uncertain) input parameters of the AGB models such as the 
mass loss prescription also influence the theoretical predictions and 
are important for nucleosynthesis purposes (see e.g., Forestini \& Charbonnel 1997).

The present simple analysis precludes us from deriving further conclusions that 
require sophisticated and complete stellar evolution computations. 
During the AGB phase the surface and wind composition of the stars 
depend indeed on the subbtle competition between HBB and third dredge-up. 
As a matter of fact detailed studied based on custom-made AGB models 
pointed out several severe drawbacks of the AGB pollution scenario 
(Ventura el al. 2001, 2002; Denissenkov \& Herwig 2003; Karakas \& Lattanzio 
2003; Herwig 2004a, b; Ventura \& D'Antona 2005a, b, c; Karakas et al. 2006).
These difficulties come from the fact that the third dredge-up does contaminate 
the AGB envelope with the products of helium burning ($^{16}$O and $^{25,26}$Mg 
in particular) and creates abundance patterns in conflict with the ones observed 
(see e.g., Fenner et al. 2004, Charbonnel 2005, and PC06). 
In particular, the sum C+N+O does not remain constant in AGB processed 
material\footnote{Note however that Ventura \& D'Antona (2005a) have managed
to keep the total C+N+O abundance within a factor of $\sim$ 3 in their AGB models
computed with the FST model for convection and with some ``extra-mixing" 
at the base of the stellar convective envelope.}, 
in contrast to observational requirements: whenever the relevant data are 
available simultaneously, the sum C+N+O (as well as Mg+Al) is indeed found 
to be constant from star-to-star within the observational errors
(Dickens et al. 1991, Ivans et al. 1999).
PC06 discuss other shortcomings of the AGB scenario, 
related e.g. to the peculiar initial mass function it requires; in addition, 
they underline the fact that the AGB scenario gives no satisfactory answers
as to the role of stars more massive and less massive than the presumed 
polluters and does not provide a mechanism to trigger the formation of the 
chemically peculiar low-mass stars that we observe today in GCs.

\subsection{Massive stars}

Finally we see from Fig.~8 that the central temperature of massive main sequence
stars (i.e., with an initial mass slightly higher than 40~M$_{\odot}$)
reaches the values required to produce the observed abundance anomalies. 
In Decressin et al.~(2007; see also PC06) we suggest that fast rotation near 
the breakup allows massive stars to eject their H-burning products through slow winds.  
The theoretical surface and wind composition of fast rotating massive stars 
presents thus the H-burning signatures observed in GC stars. 
A difficulty within this framework concerns the destruction 
of $^{24}$Mg and the related amplitude of the Mg-Al anticorrelation; 
they require indeed a large increase of the $^{24}$Mg(p,$\gamma$) 
reaction rate around 50~MK with respect to the published 
values in order to be fully reproduced in complete stellar evolution models.
This is due to the fact that the temperature necessary to efficiently destroy 
$^{24}$Mg is reached in the core of massive stars only at the very end of 
their main sequence evolution (see Fig.~7 of Decressin et al. 2007).
As mentioned in \S~3.2 the $^{24}$Mg(p,$\gamma)^{25}$Al reaction has a 
small rate error in the temperature range we consider (see Powell et al. 1999). Note, 
however that a low-lying resonance can not be totally excluded (although, based 
on the known level structure in these nuclei, no resonance is predicted).

For the moment we stress that the Winds of Fast Rotating Massive Stars 
scenario presents a very interesting framework for the understanding of the 
self-enrichment of GCs. Fast rotation\footnote{By fast rotation we mean  
an initial rotation velocity such that the star reaches the so-called 
critical velocity early on the main sequence. As an example, this
corresponds to 400~km s$^{-1}$ on the zero age main sequence for 
a 60~M${\odot}$ star.} loads the surface layers as well as 
the stellar wind with the appropriate H-burning products. It helps removing 
that material from the stellar surface through low-velocity stellar winds 
which can be easily kept within the GC. We note in addition that the abundance 
patterns are well reproduced when assuming mixing of the massive star ejecta 
with pristine gas as required by the extreme observed value 
of Li in Na-rich stars. Last but not least, massive stars can, 
through the SN shocks or the ionisation front they produce, trigger star formation 
in their vicinity (PC06). However, in that case also (as in the case of the AGBs), 
the IMF of the first generation stars has to be flatter than a normal IMF in order
to explain observations (PC06, Smith 2006). 

\section{Summary}

In this work, we study quantitatively the nucleosynthetic processes that may be at 
the origin of the peculiar abundance patterns observed in stars of GCs. 
Our analysis is adapted, in particular, to the case of NGC~6752, for which a large 
body of observational data is available. The data, presented in \S~2, concern 
abundances of C, N, O, Na, Mg and Al, as well as Mg isotopes, observed in about 
two dozen stars of that GC.

In order to constrain the physical conditions (i.e. the temperature range of 
the H-burning) that gave rise to the observed abundance patterns, we perform 
nucleosynthesis calculations at constant T. We adopt ``realistic" initial 
abundances (i.e. corresponding to field halo stars of similar [Fe/H]=--1.5 
as NGC~6752), a detailed nuclear reaction network and state-of-the-art 
nuclear reaction rates (presented in \S~3.1).

We find that {\it there exists a rather narrow temperature range around T$\sim$75 MK 
where the  observed extreme abundances of all elements and isotopes in NGC~6752 is 
nicely co-produced}. This happens {\it only after mixing of the nuclearly processed 
material  with $f\sim$30\% of material of pristine composition} (\S~3.3). This mixing 
factor is imposed by the fact that the observed extreme oxygen abundance in NGC~6752 
is only $\sim$5-7 times lower than the original one, and never as low as in H-burning 
regions. In fact, none of the abundances in H-burning zones corresponds to the extreme 
observed ones; the latter are recovered only after mixing with 30\% of pristine material. We note that in the framework 
of our study (which assumes no enhancement of the starting abundances for any 
of the involved nuclei, just field halo star composition) the observed \mgc \ excess 
in NGC~6752 can only be understood in terms of its production as $^{26}$Al.

We show then quantitatively how, by increasing the mixing factor $f$, one can 
naturally obtain  the full range of observed abundances in NGC~6752, i.e. the 
well-known anti-correlations between N and O, O and Na or O and Al; the behaviour 
of the Mg isotopes is also satisfactorily reproduced (\S~3.4).

Mixing of processed with pristine material can also naturally explain observed 
anticorrelations  of the fragile elements Li and F vs. Na in GCs; this is 
demonstrated quantitatively in \S~4, based on our results for Na production. 
Observations of heavy elements constrain the temperature range for H-burning 
and suggest mixing, but leave unclear the origin of the  mixed material (ISM or 
from the stellar envelopes); observations of the fragile light elements clearly 
point to ISM origin for the mixed material, since the fraction of the stellar 
envelopes with T$<$2.2~MK (necessary to preserve Li) is negligible.
We also note that our analysis supports the conclusion that NGC~6752 turnoff stars 
did undergo a uniform depletion of their original surface Li abundance similar to
that of field halo counterparts. It shows finally that Li production by the 
stars that build up the other abundance anomalies is not mandatory 
in the case of NGC~6752.

Finally, in \S~5 we discuss the temperature constraint found in our analysis, 
which we consider to be quite robust, in view of the sensitivity of nuclear 
reaction rates to temperature. We note that, in the case of convective regions
(such as AGB envelpes and massive star cores) the temperature we determined through our analysis
is only an effective temperature (corresponding to an average reaction rate over the whole convective region)
and provides a lower limit to the actual maximum temperature of the convective zone. We argue that, in view of the
temperature dependence of the reaction rates, the {\it maximal temperature of the convective zone
should be 10-15\% larger than the one-zone temperature of our models}, i.e. in the 80-85 MK range.
We find that such temperatures are encountered 
in the two main candidate polluters, namely massive AGB stars (with HBB) and the
most massive stars (above 40 \ms) on the main sequence. We note, however, that the precise T values are 
much more model dependent in the former case than in the latter.

\begin{acknowledgements}
We thank the anonymous referee for constructive comments. 
We are indebted to T.Decressin, L.Siess and P.Ventura for providing us 
with some informations on their stellar models. 
\end{acknowledgements}

{}

\end{document}